\documentclass{midl} 

\usepackage{mwe} 
\jmlrvolume{-- Under Review}
\jmlryear{2026}
\jmlrworkshop{Full Paper -- Submitted 2026}

\title[Physics-Guided Dual-Domain Plug-and-Play ADMM for Low-Dose CT Reconstruction]{Physics-Guided Dual-Domain Plug-and-Play ADMM for Low-Dose CT Reconstruction}

\midlauthor{\Name{Sayantan Dutta\nametag{$^{1 \ast}$}} \orcid{0000-0002-8187-4193} \Email{sayantan.dutta1@gehealthcare.com}\\
\Name{Sudhanya Chatterjee\nametag{$^{1}$}} \Email{sudhanya.chatterjee@gehealthcare.com}\\
\Name{Ashwini Galande\nametag{$^{1}$}} \Email{ashwini.galande@gehealthcare.com}\\
\Name{K. S. Shriram\nametag{$^{1}$}} \Email{ks.shriram@gehealthcare.com}\\
\Name{Bipul Das\nametag{$^{1}$}} \Email{bipul.das@gehealthcare.com}\\
~~\\
\addr $^{1}$ Science and Technology Organization, GE HealthCare, Bangalore 560066, Karnataka, India.
\footnotetext[1]{Corresponding Author: Sayantan Dutta}
}

\begin{document}

\maketitle
\begin{abstract}
Ultra-low-dose CT (ULDCT) imaging can greatly reduce patient radiation exposure, but the resulting scans suffer from severe structured and random noise that degrades image quality. To address this challenge, we propose a novel Plug-and-Play model-based iterative reconstruction framework (PnP-MBIR) that integrates a deep convolutional denoiser trained in a 2-stage self-supervised Noise-to-Noise (N2N) scheme. The method alternates between enforcing sinogram-domain data fidelity and applying the learned image-domain denoiser within an optimization, enabling artifact suppression while maintaining anatomical structure. The 2-stage protocol enables fully self-supervised training from noisy data, followed by high-dose fine-tuning, ensuring the denoiser's robustness in the ultra-low-dose regime. Our method enables high-quality reconstructions at $\sim$70–80\% lower dose levels, while maintaining diagnostic fidelity comparable to standard full-dose scans. Quantitative evaluations using Gray-Level Co-occurrence Matrix (GLCM) features — including contrast, homogeneity, entropy, and correlation — confirm that the proposed method yields superior texture consistency and detail preservation over standalone deep learning and supervised PnP baselines. Qualitative and quantitative results on both simulated and clinical datasets demonstrate that our framework effectively reduces streaks and structured artifacts while preserving subtle tissue contrast, making it a promising tool for ULDCT reconstruction.
\end{abstract}



\section{Introduction}
\label{sec:Introduction}


X-ray computed tomography (CT) is essential for diagnosis but exposes patients to ionizing radiation, posing long-term risks, especially for children and those needing frequent scans \cite{smith2010radiation}. Guided by the ALARA principle and initiatives like \textit{Image Gently}, minimizing exposure while preserving diagnostic quality is critical \cite{Brenner2007Computed, Olawade2025AI}. Pediatric CT has been linked to higher risks of leukemia and brain tumors \cite{Brenner2007Computed}, emphasizing the need for ultra-low-dose CT (ULDCT) protocols that improve safety without compromising utility.

However, reducing the tube current or voltage in CT comes at the cost of significantly degraded image quality. ULDCT images are affected by increased quantum noise (Poisson mottle) and severe streak artifacts from photon starvation, which obscure low-contrast features and anatomical details \cite{Yinjin2025deep}. The resulting low signal-to-noise ratio (SNR) can mask subtle pathologies, making reliable diagnosis challenging. Various reconstruction strategies have been developed to address these issues, broadly classified into (1) analytical methods \cite{yazdanpanah2016algebraic}, (2) model-based iterative reconstruction (MBIR) \cite{Sidky2008Image, Zhang2014Iterative, Kim2015Sparse}, and (3) modern learning-based techniques \cite{Hu2017Low, Xu2012Low, Jin2017deep, fu2022deep}.

Filtered back projection (FBP) is computationally efficient but highly noise-sensitive. Sinogram-domain filtering \cite{Balda2012ray, Manduca2009Projection} partly alleviates this, though excessive smoothing often erodes fine structural details \cite{Jing2006Penalized}. MBIR enhances image quality through statistical noise modeling \cite{Bouman1996A} and priors such as total variation \cite{Zhang2016Statistical}, non-local similarity \cite{Zhang2016Spectral, dutta2022novel}, dictionary learning \cite{Xu2012Low}, and low-rank constraints \cite{Cai2014Cine, dutta2021quantum}. Despite these advantages, MBIR remains computationally intensive and constrained by proprietary, vendor-specific implementations which hinders clinical adoption \cite{katsura2012model}.


Deep learning (DL) has become a powerful approach for low-dose CT (LDCT) denoising, offering fast, data-driven alternatives to traditional reconstruction. CNNs \cite{Hu2017Low}, U-Nets, and GANs \cite{Jin2017deep, Yinjin2025deep} learn nonlinear mappings from low-dose to high-dose images, capturing priors that suppress noise while preserving anatomy. Hybrid methods \cite{fu2022deep} and plug-and-play (PnP) frameworks \cite{venkatakrishnan2013plug} embed pretrained denoisers \cite{Floquet2024Automatic, dutta2024diva} within iterative schemes like ADMM \cite{Dutta2024Quantum, Dutta2025Enhancing}, enforcing data fidelity without explicit regularization \cite{dutta2022quantum}. Parameterized PnP \cite{Ye2018Deep, He2019Optimizing} and deep unfolding models \cite{Kong2022deep, Dutta2023Computed} achieve state-of-the-art LDCT performance, with studies confirming PnP's superiority in structure preservation and dose generalization \cite{Xu2022to}.
However, DL faces challenges: over-smoothing and loss of subtle textures or low-contrast lesions \cite{Hu2017Low}; limited generalization across scanners or dose levels \cite{Ye2018Deep, He2019Optimizing}; and reliance on large paired datasets, scarce in ULDCT \cite{Xu2022to}. Self-supervised approaches like Noise-to-Noise (N2N) \cite{chan2019Noise, lehtinen2018noise, Dutta2024Unsupervised} mitigate these issues, but purely self-supervised CT models remain hard to train and risk texture loss without careful regularization \cite{li2020SACNN, Wu2021Low}.
Another concern is image-domain inconsistency with measured sinograms, causing CT number errors. Structured noise-aware denoising \cite{Langoju2024Structured} and dual-domain strategies \cite{Fan2024MVMS, Wang2022IDOL} address this by enforcing measurement fidelity while preserving perceptual detail. Iterative refinement in both domains improves artifact suppression over single-domain methods \cite{Wang2022IDOL}, enhancing texture recovery, reducing structured noise, and preserving quantitative CT accuracy.

Building on these insights, we propose a DL-based MBIR framework (PnP-MBIR) that embeds a CNN denoiser as a proximal operator within an ADMM reconstruction loop for ULDCT. The algorithm alternates between data-consistent projection updates and image-domain denoising, ensuring fidelity to the measured sinogram at each iteration. Unlike black-box networks, our modular design separates physics-based forward modeling from learned priors. The denoiser is trained using a N2N strategy on simulated or real low-dose data across varying dose levels, enabling suppression of both structured and unstructured noise/artifacts without clean ground-truth references—critical for ultra-low-dose regime. This PnP architecture flexibly integrates realistic noise-trained priors, improving generalizability across dose levels and scan conditions.
Validated on ULDCT datasets, our method achieves superior image quality: higher entropy in uniform regions (avoiding over-smoothing), improved soft-tissue contrast-to-noise ratio, and sharper anatomical edges, yielding better visibility of low-contrast lesions and crisper anatomical boundaries compared to FBP and standalone denoising. Key contributions include:

\vspace{-2mm}
\begin{itemize}

\item \emph{Dual-Domain PnP Recon:} We propose a novel MBIR framework that incorporates deep priors in the image domain while enforcing sinogram-domain data consistency. Unlike purely image-based methods, our approach performs corrective updates in projection space at each iteration, reducing artifacts and ensuring CT number accuracy.

\vspace{-2mm}
\item \emph{Noise-to-Noise (N2N) Deep Prior:} We utilize a N2N-trained CNN denoiser as the pnP prior, trained entirely on noisy ULDCT data without requiring ground-truth high-dose images. This self-supervised setup captures real-world noise statistics (e.g., Poisson and electronic noise) of ULDCT, enhancing training practicality and robustness.

\vspace{-2mm}
\item \emph{Dose-Level Generalization:} Our 2-stage N2N training spans a wide range of dose levels (ultra-low to standard), enabling the network to learn a clean image manifold from artifact-rich pairs. This broad training improves generalization across scan conditions. To prevent potential hallucinations from unspecific mappings, the sinogram-domain correction step in our PnP framework anchors reconstructions to measured data.

\vspace{-2mm}
\item \emph{Improved Image Quality:} On ULDCT datasets (simulated and real), our method consistently outperforms FBP and baseline denoisers in both visual and quantitative evaluations. It yields higher entropy (indicating natural textures), improved soft-tissue contrast, and sharper anatomical boundaries, all while preserving attenuation accuracy and avoiding over-smoothing.

\end{itemize}


\vspace{-6mm}
\section{Methodology}
\label{sec:Methodology}

\subsection{Problem Formulation}
\label{sec:problem}

In a fan-beam CT acquisition, the measured sinogram $\mathbf{y} \in \mathbb{R}^M$ is related to the object image $\mathbf{x} \in \mathbb{R}^N$ via a forward model involving the system matrix $\mathbf{A} \in \mathbb{R}^{M\times N}$, representing the discrete fan-beam projection geometry. Under ultra-low-dose conditions, the measurement noise is no longer i.i.d.~Gaussian; instead, it becomes \textit{structured} and \textit{correlated} due to factors such as extremely low photon counts (photon starvation) and unmodeled physics (beam hardening).
These effects cause non-uniform noise variance and streak artifacts in the sinogram. In practice, noise in each ray often exhibits variance dependent on detected photon counts, and adjacent rays or views may share correlated components.


Reconstructing $\mathbf{x}$ from such noisy projections is posed as a Maximum a posteriori (MAP) estimation problem \cite{boyd2011distributed}, which seeks the image that best fits the measurements while adhering to prior expectations of image appearance. Formally, we define a cost function comprising a data fidelity term and a regularization (\textit{prior}) term:
\begin{equation}
\hat{\mathbf{x}} = \arg\min_{\mathbf{x}} \frac{1}{2}||\mathbf{A}\mathbf{x} - \mathbf{y}||_{\mathbf{W}}^2 + \lambda R(\mathbf{x})~,
\label{eq:MAP}
\end{equation}
where $||\mathbf{z}||_{\mathbf{W}}^2 = \mathbf{z}^T \mathbf{W} \mathbf{z}$ denotes a weighted $\ell_2$ norm. Here $\mathbf{W} = \Sigma^{-1}$ is a weighting (precision) matrix reflecting the noise covariance in $\mathbf{y}$. In the simplest case, $\mathbf{W}$ is taken as diagonal, $W_{ii}=1/\sigma_i^2$, to weight each projection by its noise variance, though the framework can accommodate more general $\Sigma$ capturing correlations. The term $R(\mathbf{x})$ represents the negative log-prior on $\mathbf{x}$ (i.e., regularizer), which encourages images with desired properties (e.g., smoothness or sparsity in some domain \cite{dutta2021plug}) and $\lambda$ controls the penalty weight. The goal is to find $\hat{\mathbf{x}}$ that minimizes \eqref{eq:MAP}, balancing fidelity to the data with prior-based denoising. This optimization is challenging because $N$ is large (high-dimensional images) and $R(\mathbf{x})$ is often nonconvex or implicitly defined by learned models. By selecting an appropriate regularization function, various proximal operator-based iterative schemes have been extensively studied to efficiently solve the MAP estimation problem \cite{Chan2017plug}. We address this using a model-based iterative reconstruction approach combined with a modern denoising prior, as described next.

\subsection{Plug-and-Play (PnP) ADMM Framework}
To solve \eqref{eq:MAP}, we adopt the PnP ADMM scheme, which enables the use of advanced denoisers as priors without requiring an explicit analytical form for $R(\mathbf{x})$. The PnP framework was first introduced by Venkatakrishnan et al. \cite{venkatakrishnan2013plug} and has since been successfully applied to CT reconstruction problems \cite{He2019Optimizing}. It leverages the ADMM to split the problem into manageable substeps for data fidelity and denoising. We introduce an auxiliary variable $\mathbf{v}$ to decouple the prior term from the data term:
\begin{equation}
\hat{\mathbf{x}} = \arg\min_{\mathbf{x,v}} \frac{1}{2}||\mathbf{A}\mathbf{x} - \mathbf{y}||_{\mathbf{W}}^2 + \lambda R(\mathbf{v}) \quad \text{s.t.} \quad\mathbf{x} = \mathbf{v}~.
\label{eq:split}
\end{equation}
This constrained formulation is equivalent to \eqref{eq:MAP} with the appropriate $R(\cdot)$, but allows separate handling of data fidelity and prior terms. The augmented Lagrangian for \eqref{eq:split} is:
\begin{equation}
\mathcal{L}(\mathbf{x},\mathbf{v},\mathbf{u}) = \frac{1}{2}||\mathbf{A}\mathbf{x} - \mathbf{y}||_{\mathbf{W}}^2 + \lambda  R(\mathbf{v}) + \frac{\beta}{2}||\mathbf{x} - \mathbf{v} + \mathbf{u}||_2^2~,
\label{eq:auglag}
\end{equation}
where, $\beta>0$ is the penalty parameter and $\mathbf{u}$ is the scaled dual variable (Lagrange multiplier) for the constraint $\mathbf{x}=\mathbf{v}$. Under the ADMM framework, the augmented Lagrangian in eq.~\eqref{eq:auglag} decomposes into three distinct subproblems:
\begin{subequations}
\begin{small}
\begin{align}
& \mathbf{x}^{k+1} = \arg\min_{\mathbf{x}}  \frac{1}{2}||\mathbf{A}\mathbf{x} - \mathbf{y}||_{\mathbf{W}}^2 + \frac{\beta}{2}||\mathbf{x} - \mathbf{v}^k + \mathbf{u}^k||_2^2~,
 \label{eq:x-update} \\
& \mathbf{v}^{k+1} = \arg\min_{\mathbf{v}} \lambda R(\mathbf{v}) + \frac{\beta}{2}||\mathbf{x}^{k+1} + \mathbf{u}^k - \mathbf{v}||_2^2~,
 \label{eq:v-update} \\
& \mathbf{u}^{k+1} = \mathbf{u}^k + \mathbf{x}^{k+1} - \mathbf{v}^{k+1}~.
\label{eq:dual-update}
\end{align}
\end{small}
\end{subequations}
\noindent Eq.~\eqref{eq:x-update} is a quadratic minimization that updates the image to enforce data fidelity. Setting the gradient to zero yields the normal-equation system:
\begin{equation}
\big(\mathbf{A}^T \mathbf{W} \mathbf{A} + \beta I\big)\mathbf{x}^{k+1} = \mathbf{A}^T \mathbf{W}\mathbf{y} + \beta(\mathbf{v}^k - \mathbf{u}^k)~,
\label{eq:normal}
\end{equation}
is solved for $\mathbf{x}^{k+1}$ using a Cholesky-based approximation for computational efficiency \cite{yeung2025algebraic}. When $\mathbf{W}$ includes ramp-filter weighting, $\mathbf{A}^T \mathbf{W}\mathbf{y}$ corresponds to an FBP image, which can serve as an initial guess or one-step approximation for \eqref{eq:x-update}, accelerating convergence. This physics-guided integration ensures that each update remains consistent with the measured sinogram data.


Eq.~\eqref{eq:v-update} is the prior enforcement step, which typically has no closed-form solution if $R(\mathbf{v})$ is a complex learned prior. In the PnP methodology, we replace this optimization with an equivalent denoising operation: instead of explicitly minimizing \eqref{eq:v-update}, we set
\begin{equation}
\mathbf{v}^{k+1} = \mathcal{D}_{\sigma}\big(\mathbf{x}^{k+1} + \mathbf{u}^k\big)~,
\label{eq:denoiser-update}
\end{equation}
where $\mathcal{D}_{\sigma}(\cdot)$ is a learned image denoiser (with implicit parameter $\sigma$ controlling its noise reduction strength) that acts as the proximal operator for $R$. In other words, $\mathcal{D}_{\sigma}$ plays the role of $\mathrm{prox}~{\lambda R/\beta}$, producing a denoised image $\mathbf{v}^{k+1}$ that balances staying close to $\mathbf{x}^{k+1}+\mathbf{u}^k$ with having desirable image properties encoded by the prior. The dual update \eqref{eq:dual-update} then adjusts $\mathbf{u}$ to enforce consistency between $\mathbf{x}$ and $\mathbf{v}$.

This ADMM-based splitting enables integration of advanced denoisers within a rigorous iterative framework. PnP-ADMM offers convergence guarantees under mild conditions \cite{Chan2017plug}, and our implementation exhibits stable behavior similar to prior LDCT studies \cite{He2019Optimizing}. Penalty $\beta$ and denoiser strength $\sigma$ are tuned to ensure diminishing residuals across iterations. Overall, the framework combines physics-based data consistency with learned priors for robust ULDCT reconstruction.


\subsection{Denoiser Training Protocol}
\label{sec:TrainProto}

The performance of PnP-MBIR depends on the denoiser prior $\mathcal{D}_{\sigma}$, built on a residual dense network (RDN) backbone \cite{zhang2018residual} with attention modules \cite{woo2018cbam} to emphasize informative features like edges and soft-tissue textures \cite{ma2021sinogram}. It comprises residual dense groups with interleaved attention layers and skip connections for learning residual noise maps.
We adopt an image-domain residual learning strategy where the network predicts noise rather than mapping low-dose to high-dose directly, focusing on structured artifacts with lower dynamic range and distinct patterns \cite{Wagner2023On}. Following the RED-CNN paradigm \cite{Hu2017Low}, the predicted residual is added to the input, enabling suppression of stochastic and structured noise while preserving diagnostic details.
However, training $\mathcal{D}_{\sigma}$ is challenging due to the absence of clean ground-truth CT data for low-dose scans. To overcome this, we employ a 2-stage N2N strategy inspired by Lehtinen et al. \cite{lehtinen2018noise}, tailored for ultra-low-dose conditions.

\paragraph{Stage 1 – Criss-cross N2N pre-training:}

The denoiser is first trained in a self-supervised manner using pairs of independently simulated noisy data. Two low-dose reconstructions or FBP images of the same object are used—one as input, the other as target—minimizing their discrepancy drives the network toward the underlying noise-free signal. To increase diversity, input and target roles are randomly swapped (criss-cross strategy) \cite{calvarons2021improved}, doubling the dataset and exposing the network to both noise-to-noise directions. This approach expands training data and removes the need for ground-truth high-dose images.
The networks in this stage are trained to minimize a combined mean-squared error (MSE) and structural similarity (SSIM) loss between the network output and the target noisy measurement, $\mathcal{L}_{\text{stage1}} = \mathcal{L}_{\text{MSE}} + \lambda_{\text{SSIM}} \mathcal{L}_{\text{SSIM}}$, with $\lambda_{\text{SSIM}}$ as tunable weight.
Since the true signal is the only consistent component across noise realizations, this stage promotes meaningful noise suppression. By leveraging diverse noise patterns in a criss-cross manner, the denoiser becomes robust to mild and severe noise and learns realistic LDCT artifacts (structured, textured, and correlated), unlike synthetic additive noise.

\paragraph{Stage 2 – Supervised fine-tuning:}

After pre-training, networks are fine-tuned on paired low-dose and high-dose data, predicting residual maps from low-dose inputs to high-dose outputs. Initialized with Stage~1 weights, training uses a hybrid loss combining $\mathcal{L}_{\text{MSE}}$ for general denoising with terms that improve residual accuracy. This step calibrates denoisers to dose-specific noise (correlated and structured artifacts) and corrects Stage~1 bias, mitigating over-smoothing while preserving texture learned earlier and driving outputs closer to high-dose quality in ultra-low-dose conditions.

The denoiser is optimized using a composite loss $\mathcal{L}_{\text{stage2}}$ balancing fidelity, texture realism, and soft-tissue contrast \cite{Langoju2024Structured}. To handle structured streaks from photon starvation—often misinterpreted as true features and varying with dose—additional loss terms are introduced alongside $\mathcal{L}_{\text{MSE}}$.

\emph{Entropy-based texture regularization:} This term encourages the retention of anatomical texture by maximizing output entropy, preventing over-smoothing. It is defined as the negative Shannon entropy of local patches: $\mathcal{L}_{\text{tex}} = -\sum_{i\in \Omega} H(\mathcal{D}_{\sigma}(v_{\text{low}})_i)$, where $H(\cdot)$ denotes entropy in the neighborhood of pixel $i$ \cite{Abid2025Advancements} and $\mathcal{D}_{\sigma}(v_{\text{low}})$ is the denoised low-dose image $v_\text{low}$. Minimizing $-H$ forces the network to maximize the output entropy, thus retaining a level of randomness/texture similar to that of real high-dose images instead of producing overly smooth results. This leverages the fact that entropy is a quantitative measure of image complexity and texture richness \cite{Abid2025Advancements}. By preserving entropy, the model is discouraged from eliminating fine stochastic details that characterize soft tissues, thereby maintaining realistic textures.

\emph{Soft-tissue contrast-aware loss:} To protect low-contrast anatomical boundaries (e.g., liver-kidney or gray-white matter), we employ a contrast-aware term that emphasizes soft-tissue intensity using a weighted SSIM loss: $\mathcal{L}_{\text{ca}} = \text{SSIM}(w_j\mathcal{D}_{\sigma}(v_{\text{low}})_j, w_j v_j^{\text{HD}})$, where $w_j$ increases for HU values typical of soft tissues and $v^{\text{HD}}$ is the high-dose target, placing more importance on subtle differences \cite{liu2020multi}. This helps preserve diagnostically relevant details vulnerable to suppression under pure MSE training in soft tissue – effectively acting like a perceptual loss focusing on low-contrast detail. Thus, the total image-domain loss becomes:
$\mathcal{L}_{\text{stage2}} = \mathcal{L}_{\text{MSE}} + \lambda_{\text{tex}} \mathcal{L}_{\text{tex}} + \lambda_{\text{ca}} \mathcal{L}_{\text{ca}}$,
with $\lambda_{\text{tex}}$ and $\lambda_{\text{ca}}$ balancing the trade-off between fidelity and perceptual quality. This combined objective helps the network suppress noise at different dose-levels while retaining texture and contrast in soft-tissue structures.

\section{Experimental Setup and Results}
\label{sec:Results}

\subsection{Data Generation}

We generated a diverse LDCT dataset using the CatSim X-ray CT simulation platform \cite{Zhang2024Development}. Multiple anatomically realistic anthropomorphic digital 3D phantoms (e.g., thorax, abdomen, chest, head) were scanned under varied acquisition conditions. The simulated scanner employed a fan-beam geometry with axial scanning and a detector array comprising 900 columns and 32 rows. Complete coverage of each phantom was achieved by varying the starting z-position during axial scans.
To emulate dose reduction, tube current and tube voltage were systematically varied (e.g., 20–200\,mA, 80–120\,kVp), where reduced X-ray flux resulted in increased image noise. The acquired projection data (sinograms) were reconstructed using standard FBP with a slice thickness of 0.625 mm to produce LDCT images, which served as inputs for subsequent experiments. By varying mA and kVp, the simulated dataset spans a broad range of noise levels and spectral characteristics, including correlated and structured artifacts. A high-dose scan at 800 mA was designated as the ground-truth reference.
Furthermore, a scan at 800mA serves as the ground‑truth reference. This range captures clinical low‑dose conditions while providing a high‑dose baseline for supervision. This diversity addresses the data-scarcity problem noted in DL reconstruction \cite{He2019Optimizing} and ensures that the PnP framework remains robust when deployed on real patient data.
Additionally, Appendix.~\ref{sec:appendixB} and Appendix.~\ref{sec:appendixC} detail the training setup and integration into the PnP-MBIR reconstruction framework, respectively.



\begin{figure*}[b!]
\centering
\includegraphics[width=\linewidth]{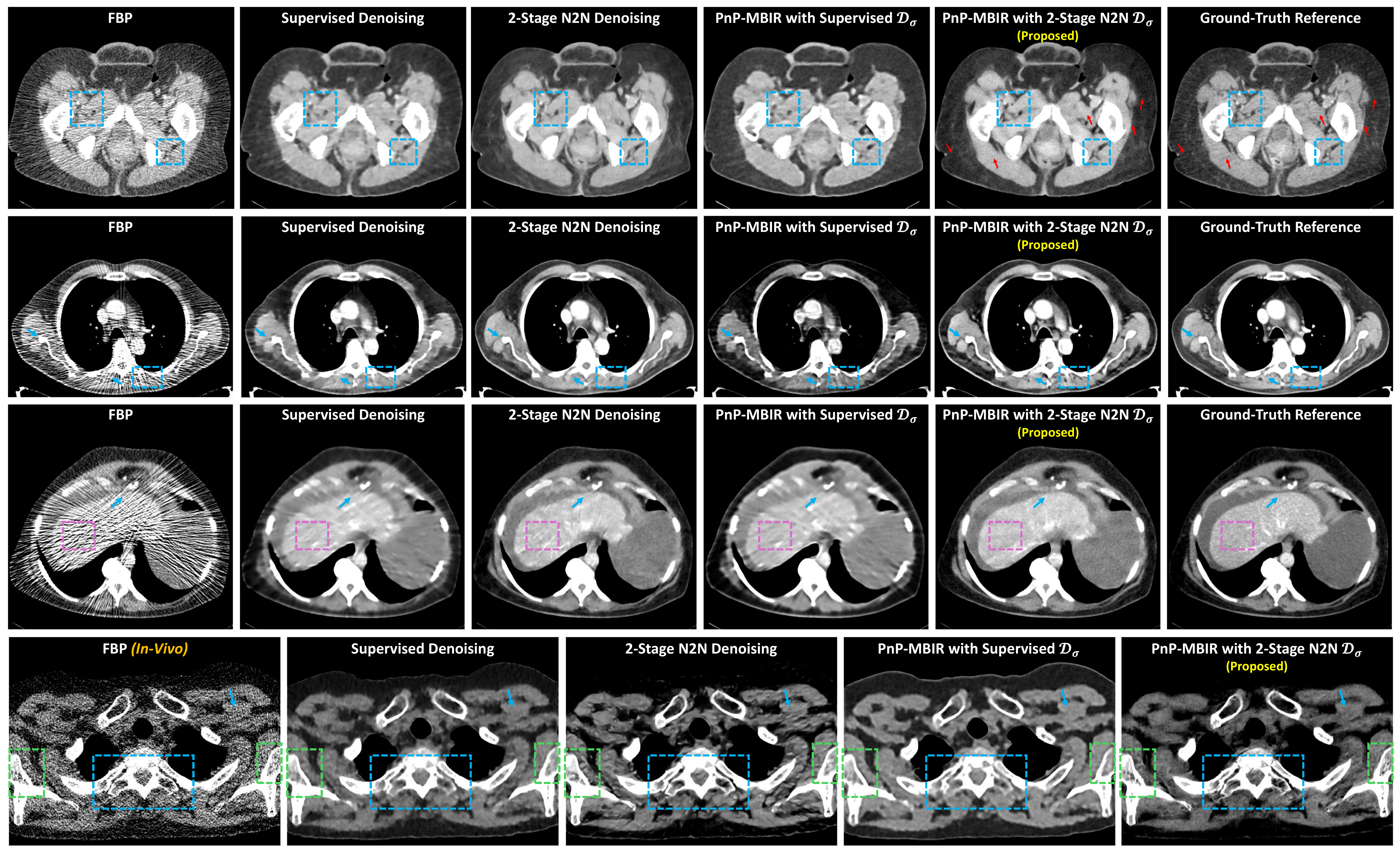}

\caption{Qualitative comparison of different reconstruction approaches. Each row shows a different anatomical image: the top three rows correspond to simulated datasets acquired at 100\,mA, 70\,mA, and 40\,mA respectively with 120\,kVp, and the bottom row shows an \textit{in-vivo} acquisition at 55\,mA with 100\,kVp. Columns (from left to right) show reconstructions from: FBP, image-domain supervised denoising, 2-stage N2N denoising, PnP-MBIR with supervised prior, and our proposed PnP-MBIR using the 2-stage (N2N) denoiser. The final column shows the ultra-high-dose ground truth image for reference (not available for the \textit{in-vivo} case). The proposed method clearly suppresses structured artifacts and preserves fine anatomical details. All images are displayed with [WL 30, WW 300].}
\vspace{-5mm}
\label{fig:qual_results}
\end{figure*}

\subsection{Results and Evaluation}

Fig.~\ref{fig:qual_results} qualitativly compares the reconstructions from FBP, supervised denoising, 2-stage N2N denoising, PnP-MBIR with a supervised prior, and our proposed PnP-MBIR with 2-stage (N2N) denoiser for dose-levels 100\,mA, 70\,mA and 40\,mA respectively in the top, second and third rows. Finally a \textit{in-vivo} data, acquired with 55\,mA and 100\,kVp, are presented in the bottom row. As expected, the low-dose FBP images (left column) are severely degraded by noise and streak artifacts.
The standalone supervised denoiser (column 2) removes most of the random noise but leaves substantial structured streak artifacts and creates distortion and anomalies in the underlying anatomical structure (see blue arrows and blue boxes), since conventional supervised training framework cannot effectively learn to remove correlated streaks and noise.
The standalone 2-stage N2N denoiser (column 3) further suppresses structured artifacts, but it introduces noticeable blurring and loss of fine details (see blue arrows and blue boxes).
PnP reconstruction using the supervised denoiser as a prior (column 4) improves on image-domain denoising – some low-contrast features  become clearer – but still leaves visible correlated artifacts in complex regions (e.g. liver, inter-tissue edges), resulting distorted reconstructions (magenta box) and loss of fine details (blue box).
In contrast, the proposed PnP-MBIR reconstruction with the 2-stage N2N denoiser (column 5) yields the best visual quality: streaks and structured noise are substantially removed while high-frequency anatomical details (highlighted by blue boxes and arrows) are preserved. This output closely matches the ultra-high-dose reference (column 6) and, in practice, even exceeds it by eliminating subtle artifacts present in the ground truth reference (red arrows) due to photon starvation and attenuation in dense bone regions.
A similar pattern holds \textit{in-vivo} (bottom row): our method retains sharp bone edges and soft-tissue contrast (blue box and arrow) far better than other low-dose methods. In addition to noise and artifact reduction, our method maintains the subtle contrast between soft and hard bone tissues, ensuring better visualization of intra-bone structures (green boxes).
Additional zoomed-in sections of the reconstructed images are provided in Appendix~\ref{sec:appendixD} for a comprehensive visual analysis.

Performance is quantified using texture and histogram-based metrics. We compute gray-level co-occurrence matrix (GLCM) features—contrast, entropy, energy, homogeneity, correlation, dissimilarity, and angular second moment (ASM)—to capture textural content and smoothness. Additionally, Earth Mover's Distance (EMD) is calculated between reconstructed and ground-truth intensity histograms. All evaluations are performed on simulated cases with known ground truth under consistent hyperparameters: GLCM with distance = 1, angles = $\{0, \frac{\pi}{4}, \frac{\pi}{2}, \frac{3\pi}{4}\}$, and 256 gray levels; EMD using 256-bin normalized histograms.

\begin{figure}[t]
\centering
\includegraphics[width=0.7\linewidth]{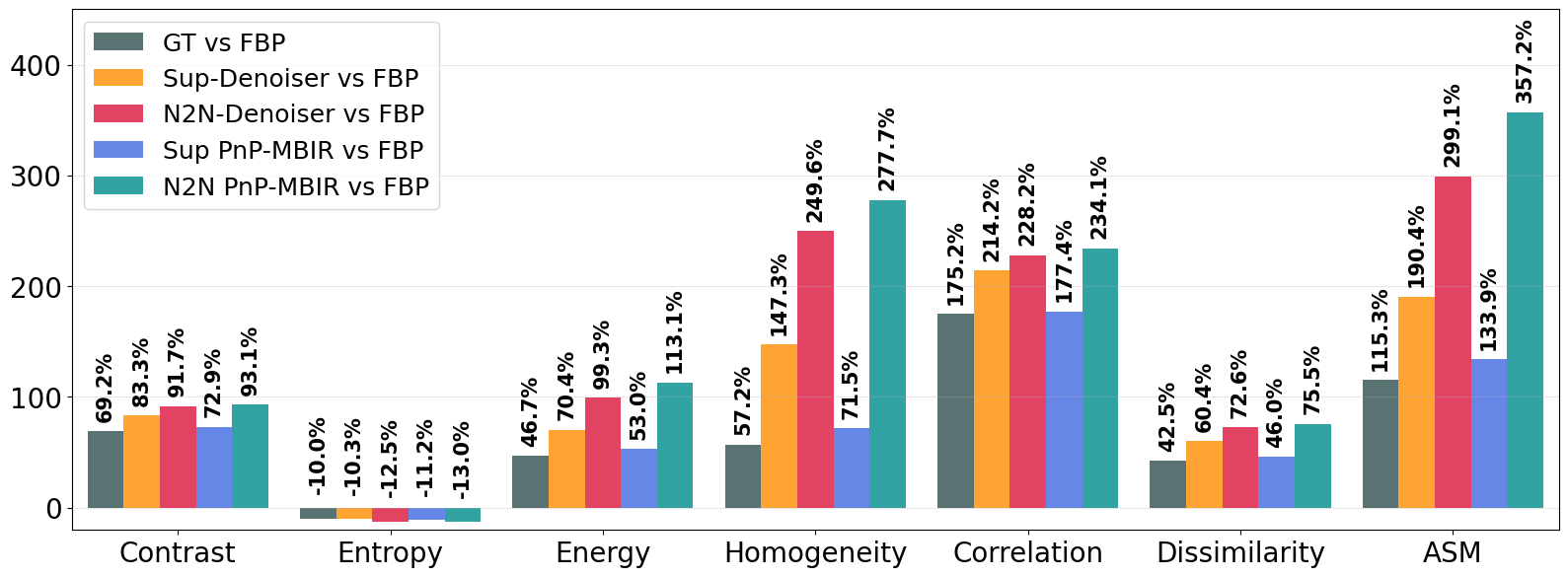}
\includegraphics[width=0.7\linewidth]{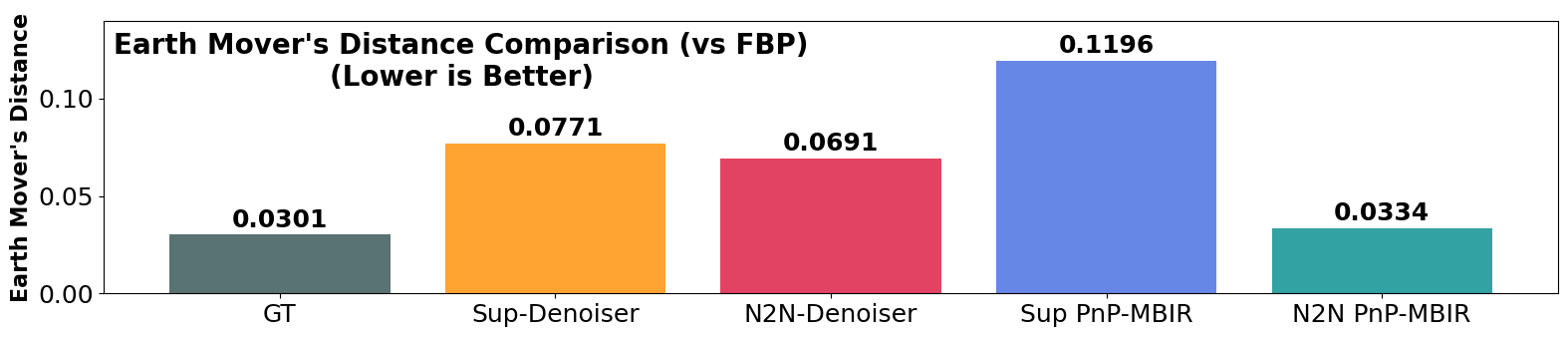}

\caption{Quantitative texture analysis using GLCM and Earth Mover’s Distance (EMD). The top plot shows relative improvement (percentage change) in key GLCM features—contrast, energy, homogeneity, correlation, dissimilarity, angular second moment (ASM), and entropy—compared to FBP. The bottom plot reports EMD values indicating distributional similarity between each reconstruction and the ground truth. Our proposed PnP-MBIR method with 2-stage N2N denoiser consistently yields the best or near-best performance across all metrics.}
\vspace{-5mm}
\label{fig:GLCM_results}
\end{figure}

The quantitative GLCM texture analysis confirms these observations. The FBP baseline has the highest contrast (1314.19) and lowest texture uniformity (energy 0.0162, homogeneity 0.0429), reflecting severe noise and artifacts. The relative changes in the GLCM measures against the FBP are summarized in Fig.~\ref{fig:GLCM_results}.
Compared to FBP, all methods significantly reduce contrast and increase uniformity. The ground-truth high-dose image already reduces contrast by 69.2\% (to 405.08) and boosts energy by 46.7\%. Our PnP-MBIR with 2-stage N2N denoiser achieves the largest gains: 93.1\% contrast reduction (to 356.08), 113.1\% increase in energy (to 0.0248), and 277.7\% increase in homogeneity (to 0.0736). Dissimilarity (texture variation) is similarly minimized (14.76, 75.5\% improvement), and the ASM (texture order) is maximized (0.0006, 357.2\% improvement). All reconstruction methods yield high GLCM correlation ($\geq$0.75), indicating good structural fidelity, with our method achieving 0.7534 (234.1\% improvement over FBP). Entropy (information content) is reduced in all cases (by 10–13\%), with our proposed PnP-MBIR with N2N reaching the largest drop, suggesting effective noise removal without loss of detail. Finally, the EMD from the FBP reference is lowest for our PnP-MBIR (0.0334), indicating its intensity distribution most closely matches the reference. In summary, our 2-stage N2N PnP-MBIR reconstruction attains the most uniform textures, minimal noise/artifacts, and highest structural similarity among the tested methods.

\section{Discussion and Conclusion}
\label{sec:Discussion}

The GLCM texture analysis and qualitative comparisons together demonstrate the clear superiority of the proposed 2-stage PnP-MBIR framework. Integrating a physics-based iterative reconstruction with our learned denoiser yields consistently better image quality than either approach alone. On average, the proposed PnP-MBIR N2N method achieved the largest improvements across all texture metrics (an average of $\sim$162.5\% improvement over FBP baseline, Fig.~\ref{fig:GLCM_results}). In particular, it achieved 93.1\% noise (contrast) reduction, 113.1\% energy increase, and a remarkable 357.2\% increase in ASM, far surpassing the gains of standalone denoisers. These results underscore the synergy of model-based reconstruction and learned priors: the iterative use of the learned denoiser exploits knowledge of the CT physics to remove noise beyond what image-domain denoising can achieve. This is consistent with prior findings that PnP-MBIR can significantly reduce noise and artifacts compared to standard MBIR, and better preserve texture than deep denoising alone \cite{Xu2022to}.

Texture and structure preservation are also significantly improved. All compared methods yield high correlation ($>$0.75, Fig.~\ref{fig:GLCM_results}) with the reference, indicating that anatomical structures are largely preserved after denoising. Notably, the PnP framework particularly enhances texture fidelity: as observed by Xu et al. \cite{Xu2022to}, PnP reconstructions preserve native image texture better than standalone CNN denoising. In our results, the PnP-MBIR with 2-stage N2N output maintains subtle soft-tissue contrast and sharp bone boundaries that are blurred in other outputs (shown with blue and green boxes in Fig.~\ref{fig:qual_results}), which is reflected in its high homogeneity and low dissimilarity. The consistent entropy reduction (around 10–13\%, Fig.~\ref{fig:GLCM_results}) across methods implies that the denoisers remove random noise without discarding diagnostically relevant information. Moreover, the proposed PnP-MBIR N2N method exhibits the lowest EMD to the high-dose reference, indicating that its intensity histogram and texture patterns most closely match the ``ideal" image. Together, the quantitative metrics and visual evidence show that our approach yields statistically cleaner, more uniform images while retaining clinically important features.

These improvements have direct clinical implications. Enhanced contrast and uniformity in ultra-low-dose reconstructions can improve the detectability of subtle lesions and anatomical structures, potentially enabling further dose reduction in screening protocols. As a result, the proposed PnP-MBIR method substantially reduces noise and streak artifacts, enabling high-quality reconstructions at $\sim$70–80\% lower radiation dose that are comparable to standard full-dose images. Moreover, our method's ability to preserve soft-tissue contrast and bone detail (e.g. subtle intra-bone structures) should increase radiologist confidence in interpretation. Interestingly, our reconstruction even outperform the nominal ``ground truth" high-dose scan in texture metrics (shown with red arrows in Fig.~\ref{fig:qual_results}); this suggests that learned priors can partly overcome limitations of the reference data (e.g. photon starvation artifacts) and achieve more diagnostically valuable images.

Despite its strong performance, the proposed 2-stage PnP-MBIR framework has some limitations. While it significantly enhances image quality over conventional FBP and standalone denoising, the spatial resolution of the reconstructed images remains bounded by the fidelity of the forward model. Future work could explore integrating resolution-enhancing mechanisms directly within the MBIR forward operator to further improve fine structural recovery. Additionally, the current approach leverages deep priors only in the image domain. Extending the framework to incorporate learned priors in the sinogram domain—trained using the same two-stage N2N strategy—could enable more powerful regularization by jointly exploiting complementary information from both domains. A modified dual-domain PnP-MBIR architecture, wherein denoisers operate in parallel on both the sinogram and image spaces, holds promise for further reducing artifacts and enhancing texture fidelity.

In conclusion, the proposed dual-domain PnP-MBIR framework offers a robust and generalized framework for ULDCT reconstructions. By embedding a self-supervised N2N deep denoiser as a prior within an ADMM-based iterative reconstruction, our method achieves substantial suppression of both stochastic noise and structured noise and artifacts while preserving soft-tissue contrast and fine anatomical detail.
Experimental results on both simulated phantoms and \textit{in-vivo} patient data demonstrate that the reconstructed images exhibit markedly improved texture quality across GLCM-based metrics, significantly outperforming conventional FBP and image-domain DL approaches in preserving anatomical detail and reducing structured noise.
Qualitatively, the approach effectively eliminates streaks and mottling noise without blurring edges or subtle features, closely matching the appearance of full-dose CT. These findings demonstrate that the 2-stage N2N-trained PnP prior can reliably enable ULDCT by restoring diagnostic image quality without sacrificing clinically relevant structures.



\bibliography{bib_MBIR_PnP_ADMM_MIDL}

\appendix

\newpage

\begin{figure}[b!]
\centering
\includegraphics[width=\linewidth]{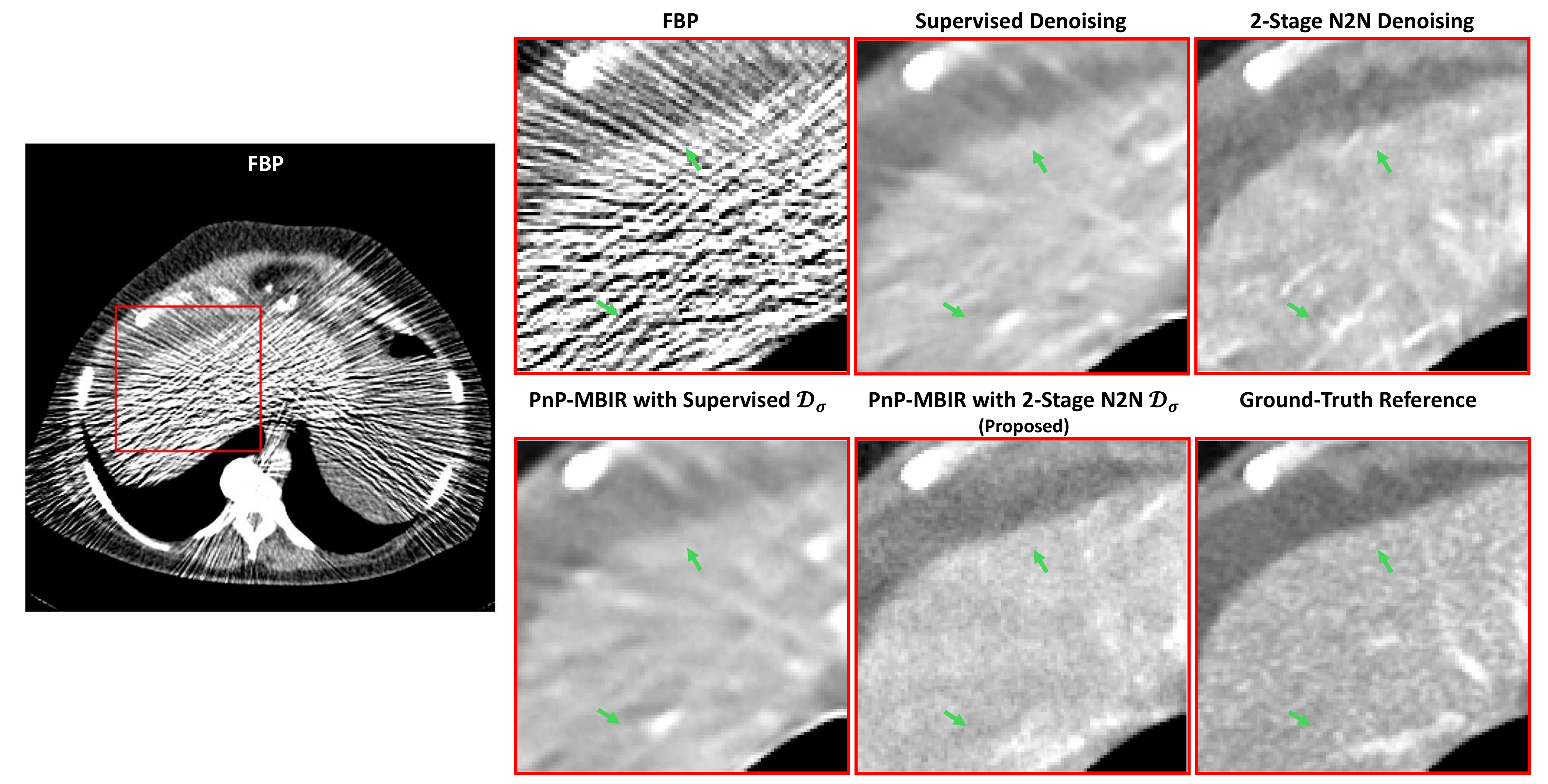}
\caption{Qualitative comparison of reconstruction methods. The left column shows a simulated FBP image acquired at 40\,mA and 120\,kVp. Zoomed-in regions (highlighted in red) from the reconstructed outputs are presented for: FBP, image-domain supervised denoising, 2-stage N2N denoising, PnP-MBIR with supervised prior, our proposed PnP-MBIR using the 2-stage (N2N) denoiser, and the ultra-high-dose ground truth image for reference. The proposed approach effectively suppresses structured artifacts while preserving fine anatomical details and soft tissue contrast. All images are displayed with [WL 30, WW 300].}
\label{fig:qual_results1}
\end{figure}

\begin{figure}[b!]
\centering
\includegraphics[width=\linewidth]{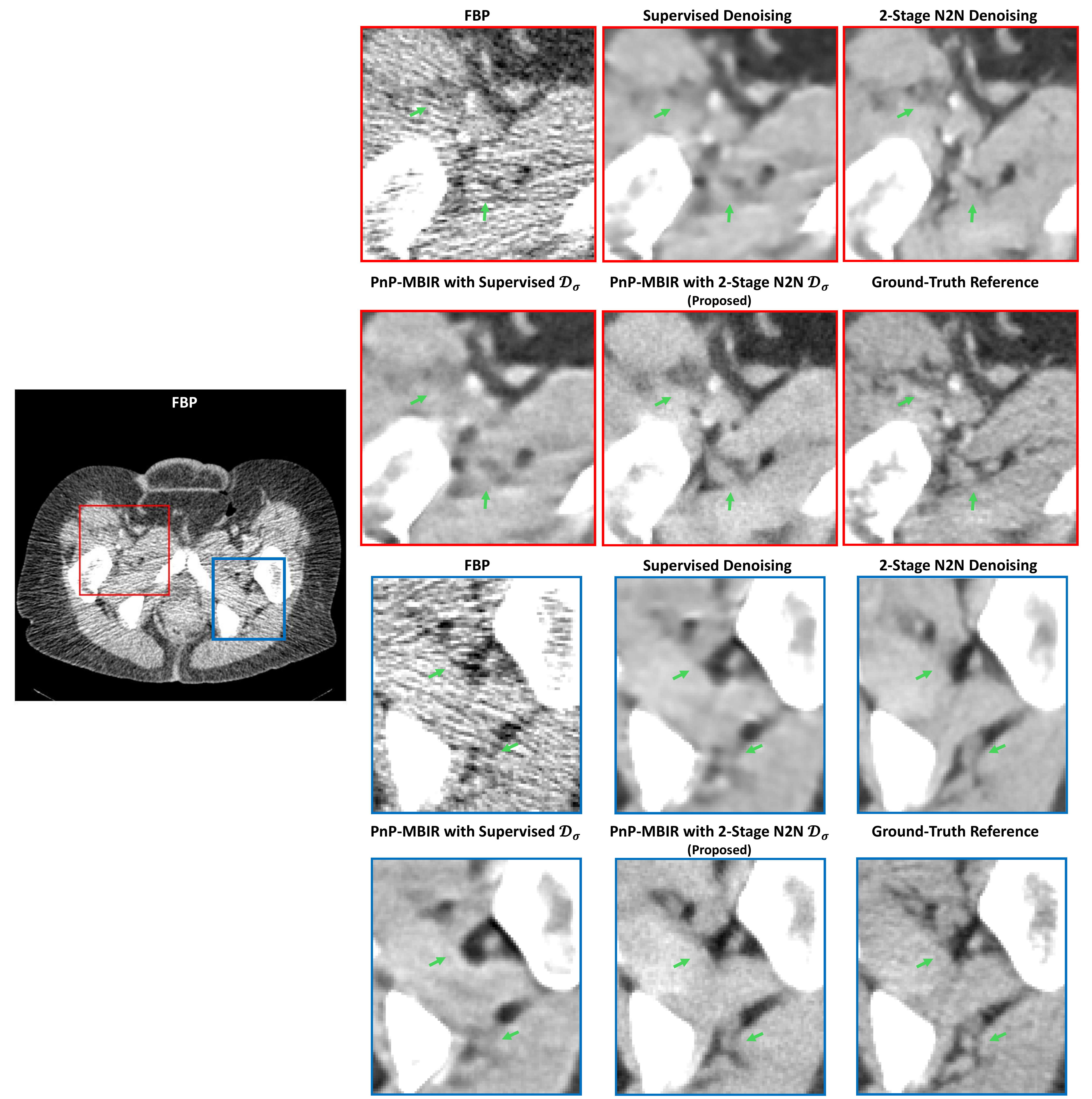}
\vspace{3mm}
\caption{Qualitative comparison of reconstruction methods. The left column shows a simulated FBP image acquired at 100,mA and 120\,kVp. Zoomed-in regions (highlighted in red and blue) from the reconstructed outputs are presented for: FBP, image-domain supervised denoising, 2-stage N2N denoising, PnP-MBIR with supervised prior, our proposed PnP-MBIR using the 2-stage (N2N) denoiser, and the ultra-high-dose ground truth image for reference. The proposed approach effectively suppresses structured artifacts while preserving fine anatomical details and soft tissue contrast. All images are displayed with [WL 30, WW 300].}
\label{fig:qual_results2}
\end{figure}

\begin{figure}[h!]
\centering
\includegraphics[width=\linewidth]{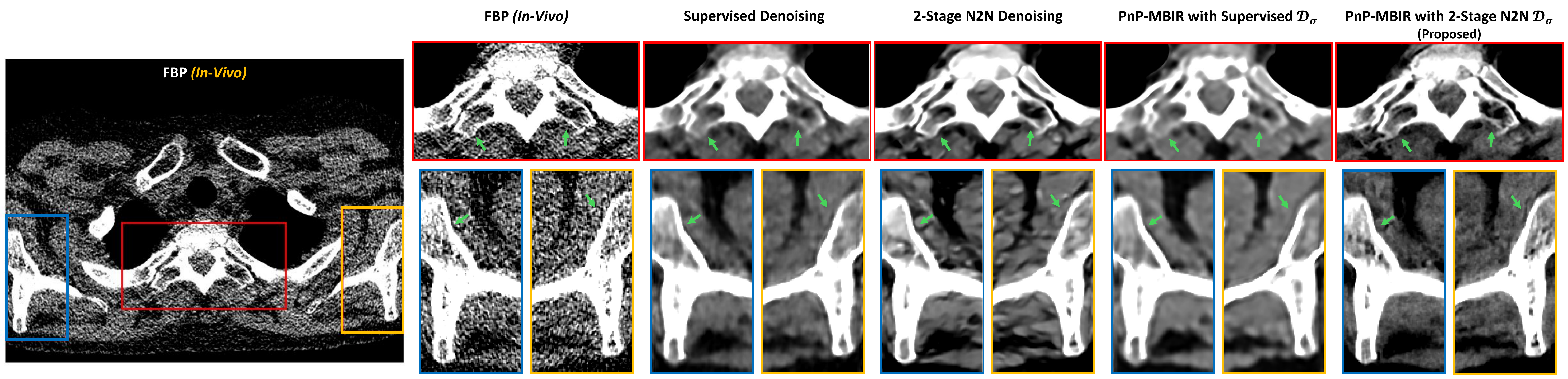}

\caption{Qualitative comparison of reconstruction methods on an \textit{in-vivo} acquisition. The left column shows an \textit{in-vivo} scan acquired at 55\,mA and 100\,kVp. Zoomed-in regions (highlighted in red, blue, and yellow) from the reconstructed outputs are presented for: FBP, image-domain supervised denoising, 2-stage N2N denoising, PnP-MBIR with supervised prior, and our proposed PnP-MBIR using the 2-stage (N2N) denoiser. The proposed approach effectively suppresses structured artifacts while maintaining subtle contrast between soft and hard bone tissues, enabling improved visualization of intra-bone structures. All images are displayed with [WL 30, WW 300].}
\label{fig:qual_results3}
\end{figure}

\section{Training Setup and Implementation}
\label{sec:appendixB}

Our image-domain denoiser is trained with $512\times512$ full resolution images.
Training uses Adam optimization (initial learning rate $1\times10^{-4}$ with a decay factor 0.99), and typical batch sizes of 8 for 500 epochs. The full-resolution data with five random augmentation followed by the 2-stage N2N training strategy make the denoisers more rebust for the ultra-low-dose regime, perticularly to learn the structured artifacts. The network is implemented in PyTorch and trained on an NVIDIA RTX 600 Ada generation cluster.
Furthermore, for compreeansive analysis of the proposed self-supervised training protocol (2-stage N2N, as described in Section~\ref{sec:TrainProto}), we trained the image-based denoisers supervised way (trained on low-dose and high-dose data) under the same dataset and training setup.

\section{Integration into PnP-MBIR Reconstruction}
\label{sec:appendixC}

The trained denoiser $\mathcal{D}_{\sigma}$ is integrated as a PnP prior within the ADMM framework. At each iteration of the proposed PnP-MBIR scheme, the projection data enforces the measurement fidelity and suppress noise in the Radon domain. This is followed by an image update step where $\mathcal{D}_{\sigma}$ refines the reconstruction by imposing natural image priors.
This alternating update process effectively regularizes both domains: eq.~\eqref{eq:x-update} ensures consistency with measured projections while removing projection noise, and $\mathcal{D}_{\sigma}$ (eq.~\eqref{eq:v-update}) suppresses residual image-domain artifacts. By leveraging the complementary strengths of both steps, the dual-domain PnP-MBIR algorithm achieves high-quality CT reconstructions with significantly reduced noise and preserved anatomical structures, even under ultra-low-dose conditions.
The iterative process continues until convergence is reached, as defined by the condition:
$\frac{1}{3N} (||\mathbf{x}^{k+1}-\mathbf{x}^{k}||_2^2 + ||\mathbf{v}^{k+1}-\mathbf{v}^{k}||_2^2 + ||\mathbf{u}^{k+1}-\mathbf{u}^{k}||_2^2) \leq 10^{-4}$ \cite{Chan2017plug}. Empirically, we found that the proposed model typically converges within five iterations when using $\beta = 1.2$.

\section{More Experimental Results}
\label{sec:appendixD}

Figs.~\ref{fig:qual_results1}–\ref{fig:qual_results3} present zoomed-in sections from the reconstructed images obtained using different methods for qualitative evaluation.
From the regions highlighted in Figs.~\ref{fig:qual_results1}–\ref{fig:qual_results2}, it is evident that our proposed PnP-MBIR reconstruction with the 2-stage N2N denoiser delivers superior visual quality. Specifically, streak artifacts and structured noise are significantly suppressed, while high-frequency anatomical details—indicated by green arrows—are well preserved. The resulting image closely approximates the ultra-high-dose reference, demonstrating the effectiveness of our approach.
A similar trend is observed in the \textit{in-vivo} results shown in Fig.~\ref{fig:qual_results3} (zoomed-in red, blue, and yellow regions). Our method maintains sharp bone edges and soft-tissue contrast (blue and yellow sections) far better than competing low-dose reconstruction techniques. Beyond noise and artifact reduction, the proposed approach preserves subtle contrast between soft and hard bone tissues, enabling improved visualization of intra-bone structures (highlighted by green arrows).

\end{document}